# Measurement of the coupling between applied stress and magnetism in a manganite thin film

Surendra Singh[1,2], M. R. Fitzsimmons[1], T. Lookman[1], H. Jeen[3,4], A. Biswas[3], M. A. Roldan[5] and M. Varela[4]

[1]Los Alamos National Laboratory, Los Alamos, NM 87545, USA

[2] Solid State Physics Division, Bhabha Atomic Research Center, Mumbai 400085, India

[3]Department of Physics, University of Florida, Gainesville, Fl 32611, USA

[4]Oak Ridge National Laboratory, Oak Ridge TN 37831 USA

[5]University Complutense, Madrid 28040, Spain.

*Abstract:* We measured the magnetization depth profile of a $(La_{1-x}Pr_x)_{1-y}Ca_yMnO_3$ ($x = 0.60\pm0.04$, $y = 0.20\pm0.03$) film as a function of applied bending stress using polarized neutron reflectometry. From these measurements we obtained a coupling coefficient relating strain to the depth dependent magnetization. We found application of compressive (tensile) bending stress along the magnetic easy axis increases (decreases) the magnetization of the film.

The complexity of doped manganites is a consequence of the competition between charge, spin, orbital and lattice order [1-4]. Collective interactions between these order parameters can lead to colossal magnetoresistance (CMR), metal insulator transitions (MIT) and coexistence of phases with different electronic properties, e.g., ferromagnetic metal (FMM), antiferromagnetic charge ordered insulator (COI), and paramagnetic insulator [5-8]. The behavior of a complex material can be greatly influenced by its environment, e.g., by magnetic field [8], light [9], stress [4, 10-16], disorder [15], etc. Theory suggests that the metal insulator phase fraction in $(La_{1-x}Pr_x)_{1-y}Ca_yMnO_3$ films can be tuned by strain [4]. In one report Millis *et al*. [14], found a shift of the ferromagnetic transition temperature $T_c$ of a CMR film could be 10% for 1% biaxial strain.

Multiphase coexistence in manganites [5-8] has been attributed to the influence of quenched disorder, e.g., chemical or strain non-uniformity, in the vicinity of a first-order transition [15], as well as long-range strain mediated interactions [4, 16]. For example, disorder can lead to pinning of phase boundaries [15] thereby inhibiting transformation. Alternatively, long range strain may influence the fractions of conducting and insulating phases (as for example by favoring or suppressing ferromagnetic order). Hence, the effect of stress on the transport and magnetic properties of manganites may provide insight into mechanisms underlying phase coexistence [5].

Several studies have reported that films exhibit unique electronic and magnetic properties that depend upon film thickness [11] or growth substrate [12, 13]. Often these differences are attributed to differences of epitaxial strain, though epitaxial strain and the extent of the strain field into the film are two of many structural features affected by film thickness and choice of substrate. In order to clarify the influence of strain on the electronic and magnetic properties of complex oxides, some studies have used structural phase transformation [17] or the piezoelectric property [18] of a substrate or have used mechanical jigs [19] to apply stress to a film grown on the substrate's surface. The first two techniques impose specific requirements for single crystal films limiting their use to only those films that can be epitaxially grown on substrates exhibiting piezoelectric response or structural transformations. In principle, mechanical jigs are more generally applicable, although the stress they exert (before fracture of the substrate) is relatively small.

Magnetization measurements under applied strain are necessary in order to quantify the change in relative volumes of the competing phases under strain. Here, we report measurements of the magnetization depth profile of a $(La_{1-x}Pr_x)_{1-y}Ca_yMnO_3$ ($x = 0.60 \pm 0.04$, $y = 0.20 \pm 0.03$)(LPCMO) single crystal film as a function of systematically applied *compressive and tensile* bending stress. We observed the application of bending stress affects the magnetization depth profile and the metal-insulator transition temperatures. A coupling coefficient relating magnetization to strain was obtained.

A 1 cm by 1 cm by 25-nm-thick single crystal film with the nominal composition of $(La_{1-x}Pr_x)_{1-y}Ca_yMnO_3$ ($x = 0.60$, $y = 0.33$) was epitaxially grown on a (110) $NdGaO_3$ (NGO) substrate in the step-flow-growth-mode using pulsed KrF laser (248 nm) deposition (PLD) [20]. During growth, the substrate temperature was 780°C, $O_2$ partial pressure was 130 mTorr, laser fluence was 0.5 J/cm$^2$, and the repetition rate of the pulsed laser was 5 Hz. The thickness of the substrate was 0.25 mm.

Bending stress was applied to the film using a four point bending jig (Fig. 1(a)). A four point jig applies stress uniformly over the lateral dimensions of a large sample [21]. Furthermore, since the film thickness is small compared to the substrate thickness, the stress is essentially uniform across the film's out-of-plane direction (the plane of zero stress lies halfway inside the

substrate)—this attribute distinguishes our approach from those of Refs. [17] and [18] which utilize techniques that produce strain gradients normal to the film's surface. Thus, our experiment is not sensitive to flexomagnetism as discussed in Ref. [22]. Tensile ($\varepsilon > 0$) and compressive ($\varepsilon < 0$) strain of the thin film can be realized by placing the film in contact with or opposite to the inner supports of the jig, respectively (Fig. 1(b)). The innovation of our experiment is the design of the four point bending jig that enables simultaneous measurement of the resistance (with current parallel to the sample's surface) and the neutron reflectivity of the sample, as functions of temperature, applied magnetic field and stress, from which the magnetic depth profile as functions of these environmental parameters can be obtained. The jig and experimental protocol are likely to have impact on a broad range of important and novel materials, piezomagnetic and multiferroic materials.

Bending stress was applied parallel to the magnetic easy axis of the sample, which is parallel to [1$\bar{1}$0] NGO [23]. The sample was cooled or warmed in the 6 kOe field (applied along the easy axis) at a rate of 0.4 K/min. Transport measurements for the sample with $\varepsilon = \pm\,0.011\%$ and $\varepsilon = 0\%$ (without applied stress) are shown in Fig. 1(c). The applied bending strain in the film is measured using: $\varepsilon = \frac{t_s}{R}$ [24], where $t_s$ and $R$ are the thickness of substrate and the radius of curvature of film, respectively. The radius of curvature of the sample was measured using a laser [25].

To independently examine the chemical depth profile, we measured a small portion of our sample with electron energy-loss spectroscopy (EELS) [26]. While the average composition of the film was $(La_{1-x}Pr_x)_{1-y}Ca_yMnO_3$ (x = 0.60±0.04, y = 0.20±0.03), we observed three chemically distinct regions (surface, film-bulk and film-substrate) [26]. The EELS data suggest a variation in the $Mn^{4+}$:$Mn^{3+}$ ratio along the depth of the film [23, 26], particularly near the film-substrate interface.

Polarized neutron reflectivity (PNR) measurements of the sample were obtained using the Asterix spectrometer at the Los Alamos Neutron Science Center [27]. In PNR the intensity of the specularly reflected neutron beam is compared to the intensity of the incident beam as a function of wave vector transfer, $Q$ ($= 4\pi \sin\theta/\lambda$, where, $\theta$ is angle of incidence and $\lambda$ is neutron wavelength), and neutron beam polarization. The specular reflectivity, $R$, is determined by the neutron scattering length density (SLD) depth profile, $\rho(z)$, averaged over the lateral dimensions of the sample. $\rho(z)$ consists of nuclear and magnetic SLDs such that $\rho^{\pm}(z) = \rho_n(z) \pm CM(z)$, where $C = 2.853\times10^{-9}$ Å$^{-2}$(kA/m)$^{-1}$ and $M(z)$ is the magnetization (in kA/m) depth profile [27]. The +(−) sign denotes neutron beam polarization parallel (opposite) to the applied field and corresponds to reflectivities, $R^{\pm}(Q)$. Thus, by measuring $R^+(Q)$ and $R^-(Q)$, $\rho_n(z)$ and $M(z)$ can be obtained separately. The reflectivity data were normalized to the Fresnel reflectivity ($R_F = \frac{16\pi^2}{Q^4}$) and are shown in Figs. 2 and 3 [27].

Before bending the sample, we measured the x-ray reflectivity (XRR) of the sample at room temperature and its reflectivity with polarized neutron beams at 200 K and 40 K in a 6 kOe strong field. The XRR and PNR data were analyzed using the method of Parratt [26, 28]. Using the insight provided by EELS, the chemical depth profile of the sample was represented by three chemically distinct regions. The model was fitted to the XRR data to obtain the thicknesses of each region and the length scale over which the chemical profile changed (from region to region). The chemical model of the sample was then used in a second model to obtain the magnetization depth profile [26] from the PNR data as the sample was bent.

In order to study the coupling between strain and magnetism, we carried out PNR measurements as a function of applied stress at constant temperature, $T$. We also performed measurements for constant ratios of temperature to transition temperature ($T/T_{IM} = T/T_{MI} = 0.93$) while cooling and warming across the MIT's. $T_{IM}$ (insulator to metal transition) and $T_{MI}$ (metal to insulator transition) temperatures are defined as temperatures at which the resistance changes by 95% of its maximum value during the cooling and warming cycles, respectively. We performed measurements at constant $T$ (open circles Fig. 1(c)) and $T$-ratios (open triangles in Fig. 1(c)) (temperatures normalized by the metal-insulator transition temperatures) in case stress changed $T_c$.

Fig. 2(c) and (d) show the $M(z)$ profile of the sample taken for constant $T$ (78 K) and $T/T_{IM}$ in Figs. 2(a) and (b) respectively. The $M(z)$ profiles were obtained from the PNR data by fitting a model in which only the magnetization of each region was varied [29]. Similarly while warming the sample, we measured PNR data for constant $T$ (89 K) and constant $T/T_{MI}$, and these data are shown in Fig. 3. The analysis shows compelling evidence that compressive strain increases the average magnetization while tensile strain suppresses the average magnetization. The trends are realized regardless of whether the measurements were made for constant temperature or constant ratio of temperature to metal-insulator (insulator-metal) transition temperature. Furthermore, the trends are the same regardless of the chemical composition of the region (e.g., as shown by regions I, II and III in the inset of Fig. 2(c)).

Upon application of small tensile (compressive) bending stress (i.e. $|\varepsilon|$ ~0.011%) we observed a shift in $T_{MI}$ (or $T_{IM}$) of about 3 K to lower (higher) temperature (Fig. 1(c)). The shift of the MIT to higher temperature and consequential reduction in resistance at fixed temperature upon application of compressive stress is consistent with other studies [17-19]. We also observed a ~ 20% increase of $M$ for compressive (negative) stress and a ~20% decrease of $M$ for tensile (positive) stress at $T = 78$ K (Fig. 2 (c)). The LPCMO film without applied stress exhibited a paramagnetic to ferromagnetic transition temperature of $T_c \sim 130$ K [20, 26]. If the large change of $M$ were attributed to a change of $T_c$, then $T_c$ would have changed by ~ 100 K to account for the change of $M$ (inferred from a fit of the Brillouin function to the temperature dependence of $M$ as measured with SQUID magnetometry for the LPCMO film without applied stress [26]). Such a shift is much larger than that suggested by Millis *et al*. [14]. Further, such a change of $T_c$ would be much larger than the 6 K change observed for a more highly strained (0.06%) $La_{0.7}Sr_{0.3}MnO_3$ film [18].

An alternative explanation for the change of $M$ with strain is to suppose the two are coupled. We considered a strain contribution to the usual free energy of a magnetic system including piezomagnetic and magnetoelastic contributions by: $F_c = \gamma \varepsilon M_\varepsilon^2 + \frac{A}{2}\varepsilon^2$, where $\gamma$, $\varepsilon$, $A$ and $M_\varepsilon$ are the coupling coefficient [units of N/A$^2$], strain, Young's modulus [units of N/m$^2$] and change of magnetization due to strain, respectively. Assuming the strain relaxes quickly, i.e., our measurements are taken for a system in equilibrium; we obtain a relation for the coupling constant by minimizing the $F_c$ with respect to strain, yielding: $M_\varepsilon^2 = -\frac{A}{\gamma}\varepsilon$. The total magnetization on application of strain is given by: $M^2 = M_0^2 + M_\varepsilon^2 = M_0^2 - \frac{A}{\gamma}\varepsilon$, where $M_0$ is magnetization of the system at $\varepsilon = 0$. Thus, from the slope of $M^2$ vs. $\varepsilon$, (Fig. 4) $-A/\gamma$ is obtained. We measured Young's modulus, $A$, of our film to be $A = 200$ GPa using nanoindentation [30].

The $M^2$ vs. $\varepsilon$ curve for different regions at constant $T$ and constant $T/T_{MI(IM)}$ ratios (Fig. 2 and Fig. 3) are plotted in Figs. 4(a-f). The slopes of these curves are negative; indicating that the compressive strain leads to an increase of magnetization. The $\gamma$'s for different regions are plotted in Fig. 4(g). $\gamma$ is smallest for LPCMO film-bulk (region II) compared to the surface and interface regions (regions I and III). A small value of coupling coefficient, $\gamma$, suggests a strong coupling of magnetism and strain in the sense that a small applied bending stress will produce a large change of magnetization. Thus, we find the coupling between strain and magnetism to be most pronounced for the stoichiometric composition of $(La_{1-x}Pr_x)_{1-y}Ca_yMnO_3$ (x = 0.60±0.04, y = 0.20±0.03) than for the surface or film-substrate interface.

For the film bulk (region II), we observed a coupling coefficient ranging from $\gamma$ = 0.00029(1) N/A$^2$ at $T$ = 78 and 89 K (for the cooling and warming cycles, respectively) to $\gamma$ = 0.00057(1) N/A$^2$ for constant $T$-ratio. The coupling coefficients for surface and interface regions (region I and III) are much larger ranging from 0.002(1) to 0.004(1) for the same temperatures and $T$-ratio. The coupling coefficients of regions I and III are approximately a factor of seven larger than that of region II. We attribute this difference to the difference between the chemistry (including Mn valence) and its influence on magnetization. The square of the ratio of magnetization of the film bulk to that of the surface and interface regions is about nine.

The influence of strain on the magnetic properties of manganites is usually interpreted within the framework of two competing effects: the double-exchange (DE) interaction and Jahn-Teller (JT) coupling [5]. The increase of the DE interaction and reduction of the JT distortion of the $MnO_6$ octahedra favor ferromagnetism [5]. The JT interaction can be reduced if the angle of the Mn-O-Mn bond approaches 180° [5]. Applied compressive stress/strain may be an effective method in adjusting the JT distortion through the change of the Mn-O bond length and the Mn-O-Mn angle [11].

In summary, we have measured the magnetization depth profiles across an LPCMO film as a function of applied stress. We found the strongest coupling between strain and magnetization for the bulk film with the composition of $(La_{1-x}Pr_x)_{1-y}Ca_yMnO_3$ (x = 0.60±0.04, y = 0.20±0.03). The coupling is weaker for the film's surface and film-substrate interface. The film-substrate interface is oxygen and $Mn^{4+}$ rich. In contrast to the behavior of magnetic transition elements (where compressive stress promotes d-band hybridization, thus lowering magnetization [31]), we find that compressive stress in our LPCMO film increases the average magnetization (over a broad range of stoichiometry), whereas tensile stress decreases the average magnetization. Our results also suggest that long range strain plays an important role across the MIT since applied stress changed both the MIT temperature (~ 3 K) and the magnetization (~ 20%) of the film.

We thank N. A. Mara for nanoidentation measurements on our LPCMO film. This work was supported by the Office of Basic Energy Science (BES), U.S. Department of Energy (DOE), BES-DMS funded by the DOE's Office of BES, the National Science Foundation (DMR-0804452) (HJ and AB), Materials Sciences and Engineering Division of the U.S. DOE (MV) and ERC starting Investigator Award, grant #239739 STEMOX (MAR). Los Alamos National Laboratory is operated by Los Alamos National Security LLC under DOE Contract DE-AC52-06NA25396.Research supported in part by ORNL's Shared Research Equipment (ShaRE) User Facility, which is sponsored by the Office of BES, U.S. DOE.

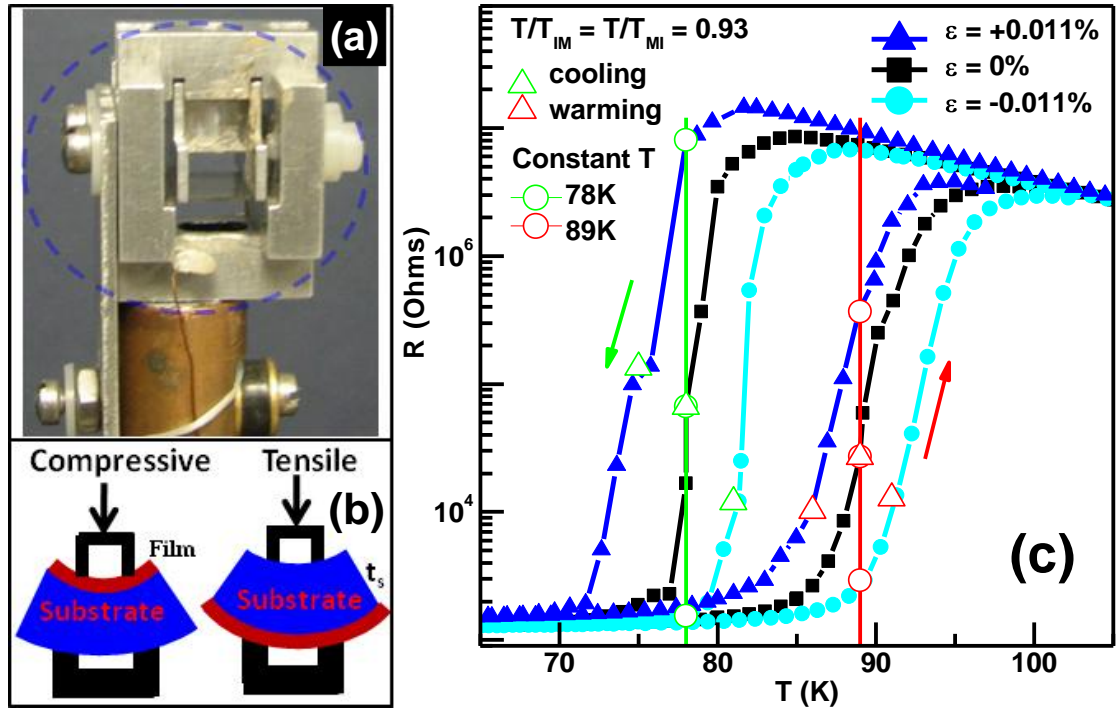

Fig. 1: End-view image of the four point jig (encircled) mounted on a cryostat (a) and side-view schematic representation of applied bending stress (b). (c): Transport measurements of the film at different applied bending stress/strain, tensile (▲), compressive (●) and no strain (■). Open circles and open triangles represents constant $T$ and constant $T$-ratio ($T/T_{IM} = T/T_{MI} = 0.93$) at which we measured PNR simultaneously.

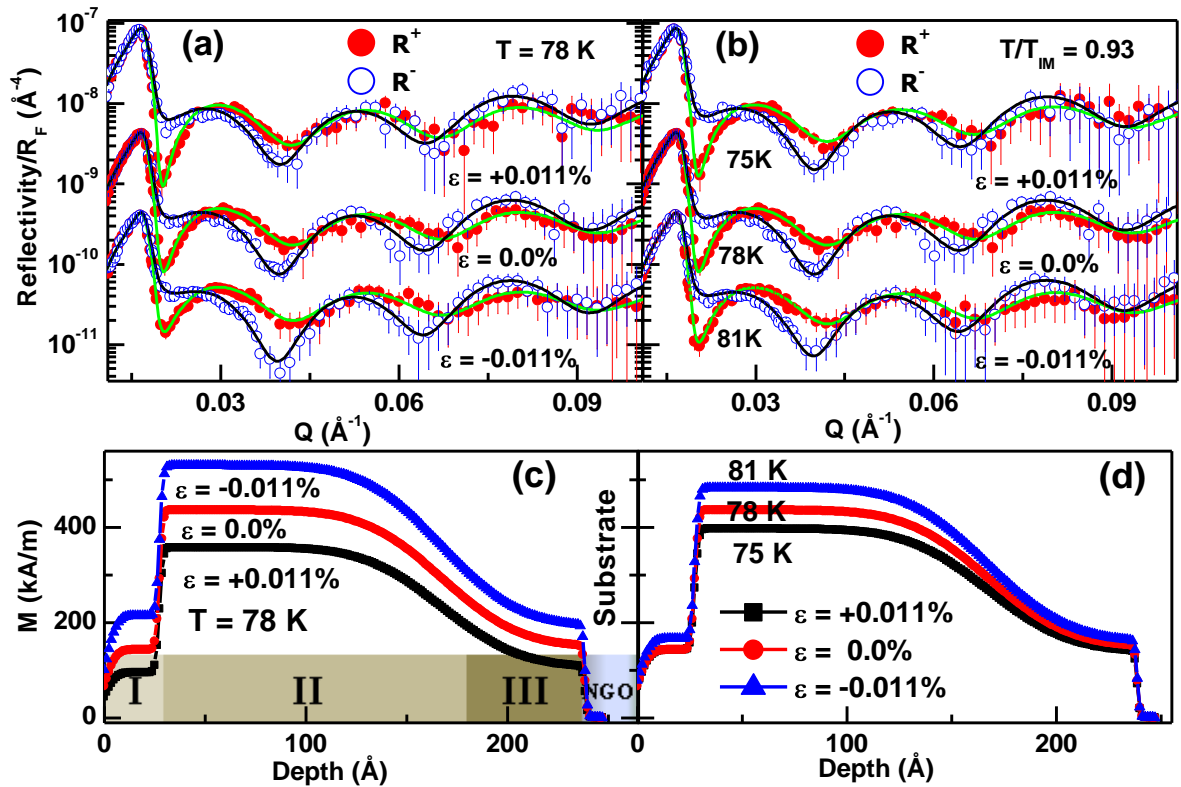

Fig. 2: PNR measurements from LPCMO film for different applied strain at constant temperature (a) and constant $T/T_{IM}$ (b) while cooling. Reflectivity data at different applied stress/strain are shifted by a factor of 5 for the sake of clarity. (c) and (d) show the magnetization ($M$) depth profile corresponding to (a) and (b) respectively.

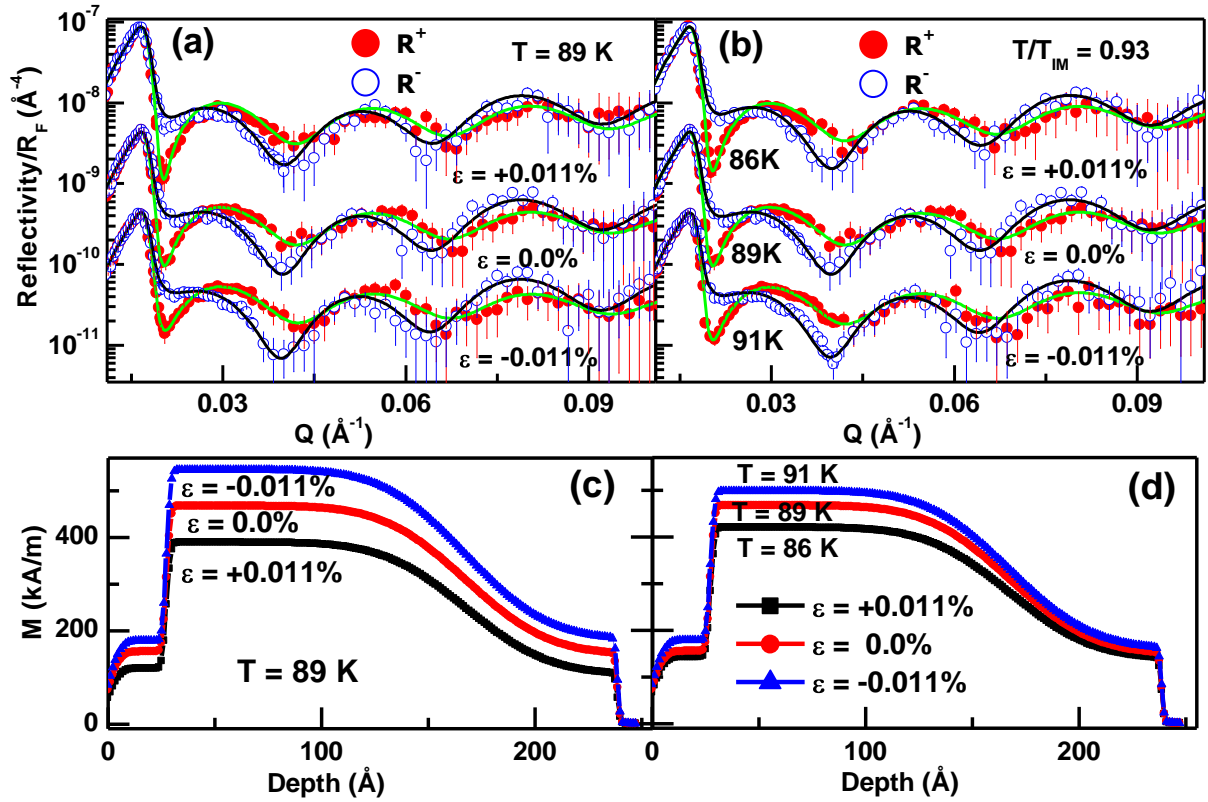

Fig. 3: PNR measurements from LPCMO film for different applied strain at constant temperature (a) and constant $T/T_{MI}$ (b) while warming. Reflectivity data at different applied stress/strain are shifted by a factor of 5 for the sake of clarity. (c) and (d) show the magnetization depth profile corresponding to (a) and (b) respectively.

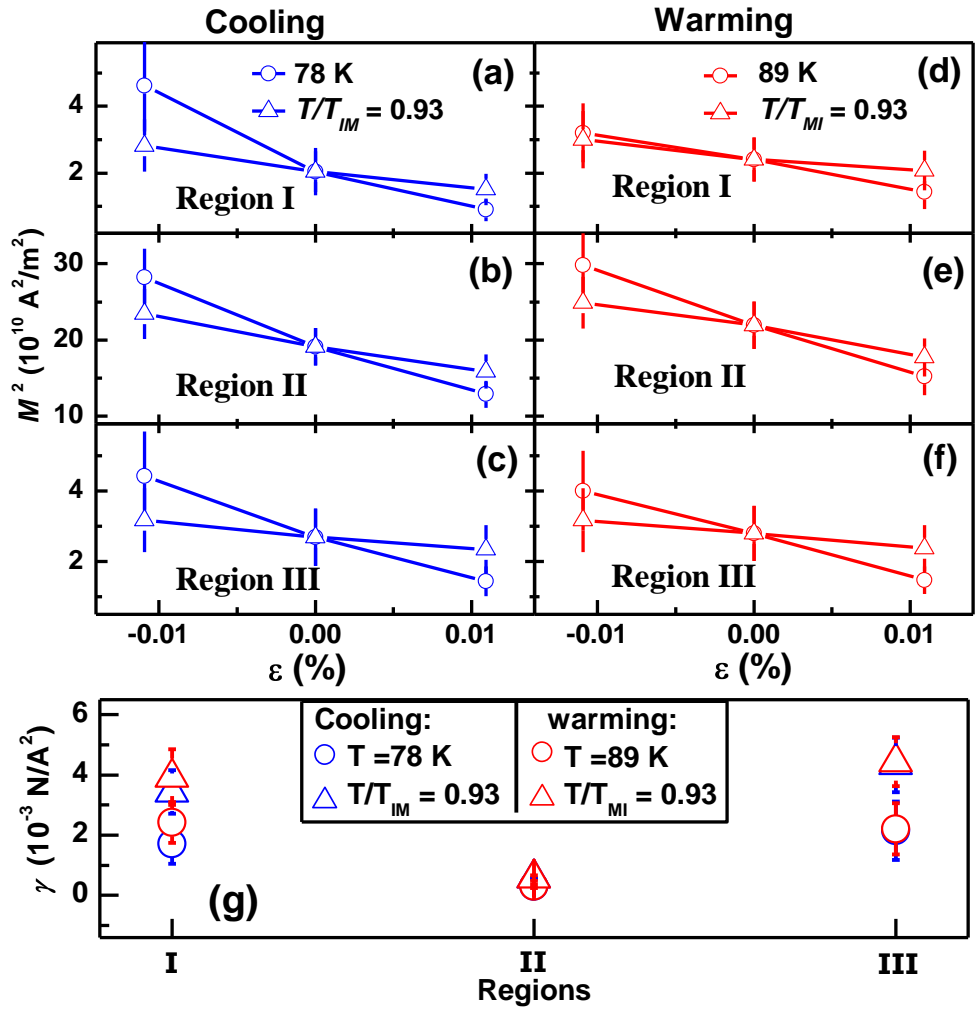

Fig. 4: (a)-(f): Variation of magnetization as a function of strain at different temperatures of cooling (left panel) and warming (right panel) for different regions of LPCMO film. The lines in the figure connect the symbols. (g): coupling coefficient, $\gamma$, for different region at different temperatures.